# The Impact of Vaccination on the Infection rate and the Severity of Covid-19

MD. Rayhan[a,b], M. Abdullah-Al-Wadud[a,c], M. Helal Uddin Ahmed[d]

---

[a] MD. Rayhan and M. Abdullah-Al-Wadud contributed equally to this work as first authors

[b] Department of Computer Science and Engineering, Brac University, Dhaka, Bangladesh.

[c] Department of Software Engineering, College of Computer and Information Sciences, King Saud University, Riyadh, Saudi Arabia. *Email addresses:* mwadud@ksu.edu.sa    *ORCID:* 0000-0001-6767-3574

[d] Department of Management Information Systems, University of Dhaka, Dhaka 1000, Dhaka, Bangladesh. *Email addresses*: helal@du.ac.bd

This study aims to statistically assess the effectiveness of vaccination against SARS-CoV-2. It is indispensable to investigate the relationship between Covid-19 deadliness and vaccination in order to study the impact of vaccine in real-world. We studied rates of infection and death due to Covid-19 in different countries with respect to their levels of vaccination. People who received the required dose of vaccination were considered as fully vaccinated in this study. Based on the percentage of fully vaccinated population, countries were categorized into several groups. Though a high-level study on the vaccine effectiveness may not provide much insight for individual level differences, a global analysis is imperative to infer the influence of vaccination as a controlling measure of the pandemic.

## Background and Methods

In this work, an exploratory data analysis of large-scale epidemiological data is carried out. Modern epidemiology entrusts investigations on massive volumes of complicated, but linkable, data to study diseases in various populations [1]. Such research has yielded insights into disease causes and outcomes, improved therapeutic targets for precision medicine, and improved disease prediction and prevention. For example, the discovery of the relationship between a province's height and water level [2] and the street water pump [3] with the outbreak of Cholera in England in 1849, which aided in the development of disease prevention techniques to bring the Cholera pandemic under control.

The worldwide health disaster caused by the Novel Coronavirus epidemic in 2020 was the world's greatest challenge in a century. Clinical mitigating measures, including vaccination approval, were put in place. In some nations, a substantial proportion of the population is fully vaccinated, whilst in others, the proportion is lower. The disparities in health outcomes across groups based on vaccination level are required to assess the clinical measure's influence. Various scientists have used statistical modeling, which has resulted in some debates on the impact of vaccination. Previous discovery found a full dose of vaccination influences infection rate decrease and significantly reduces disease lethality [4]. A recent study, on the other hand, based on infection rates over a seven-day period preceding a certain day in September 2021 and the following seven days, found that the level of vaccination is ineffective in providing defense against the virus [7]. However, the severity of the infectious condition was not investigated in this study. Our study aims to discern the effect of vaccination on the rate and severity of Covid-19 infection. We used regularly updated open-source Covid-19 research data maintained by Ritchie H et al. [5].

Considering vaccination policies and rates vary by country, we divided the 91 countries into eight categories based on the proportion of the population fully vaccinated on November 1, 2021. Immunity typically develops after 2-3 weeks of getting the full dosage [8]. As a result, in this investigation, we used a six-week period to observe vaccinations, infections, and mortality caused by the Novel Coronavirus. *Duration-1* was the last 21 days of October 2021, while *Duration-2* was the first three weeks of November 2021. During durations 1 and 2, the Delta variant was found to be present in an average of 97.32% to 98.28% of all nations [5]. We also investigated the generalizable relationship between vaccine level and Covid-19 lethality during the last three weeks of December 2021, *duration-3*, when the Omicron strain was prevalent. The severity of the disease among different groups of vaccinated people was measured based on the risk of death per confirm case. The standard metric [9] Case Fatality Ratio, *CFR*, was determined as Eq. 1 based on total confirm deaths (*CD*) and total confirmed cases (*CC*) over a three-week period.

$$CFR = (\sum CD)/(\sum CC) \ldots\ldots\ldots\ldots (1)$$

Despite the fact that there should be a delay between death cases and confirmed cases during an ongoing pandemic [9], this study used the most basic type of *CFR* computation.

# Findings

Fig. 1 depicts a bar diagram based on whether a country's infection rate increased or decreased over *duration-2* as opposed to *duration-1*. The red bars reflect the number of countries where infection climbed, whereas the blue bars represent countries where infection declined.

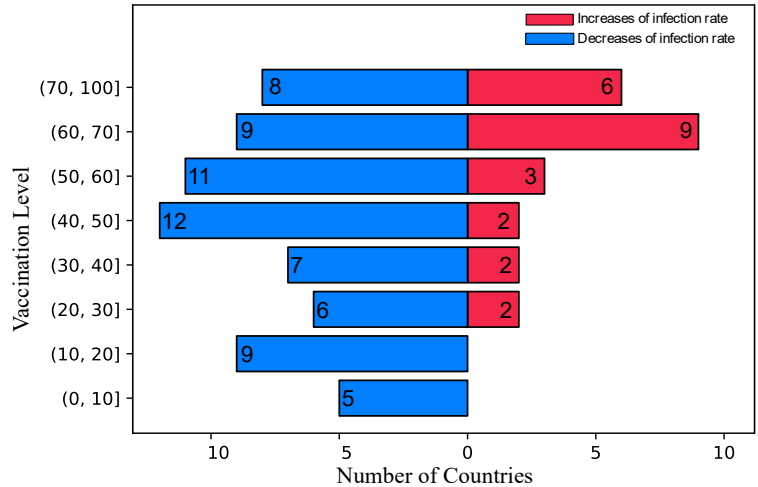

**Fig. 1** Increases and decreases of new Covid cases during the observed period in 91 countries grouped by the vaccination level

      Certainly, highly vaccinated countries had more infection cases than less vaccinated countries (Fig. 1). This pattern suggests that the level of vaccination has little effect on the infection rate. This is also analogous with the finding reported by Subramanian et al. [7], who discovered that lesser vaccinated countries had minimal transmission. However, this conjecture represents just one aspect of vaccination effectiveness studies that could be refuted in a variety of ways. For example, the extent of testing differs across countries, with some having stringent testing procedures while others do not. Partial identification approaches also demonstrate that reported cases may be significantly underestimated in comparison to actual cases [6].

      Based on the relationship between vaccination level and infection rate discussed thus far, we cannot obtain a complete picture of vaccinations' protective benefits against symptomatic infection and negative health outcomes. As a result, we focus on severity in countries classified by vaccination level in our study.

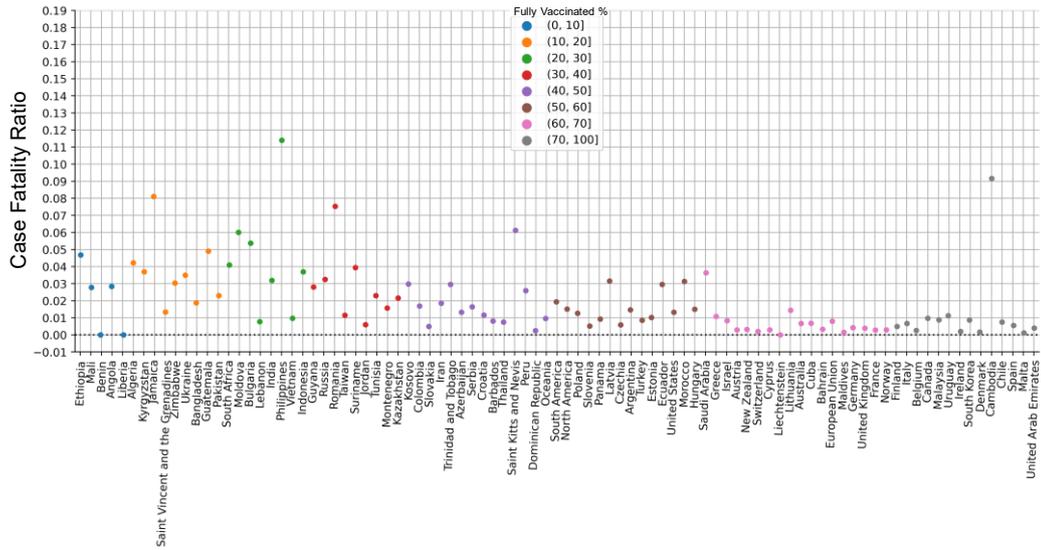

**Fig. 2** Case fatality ratios (CFR) of different countries during first three weeks of November (*duration*-2)

CFR for each country in *duration*-2 is shown in Fig. 2. It depicts that CFR is relatively lower in the countries having higher vaccination levels. It means severity of Covid-19 was lower in higher vaccinated countries. The relation between vaccination level and CFR is quantified by fitting a linear regression curve [10] in Fig. 3. It has been found that the coefficient of vaccination level on CFR was -0.0007. Furthermore, the regression model depicts that vaccination level has significant impact on CFR with an adjusted p-value<0.05. Majority of the countries having vaccination level above 65% were found case fatality ratio under 0.01. Accordingly, it is perceived that lower vaccinated countries experience more severeness of the infection than the higher vaccinated countries.

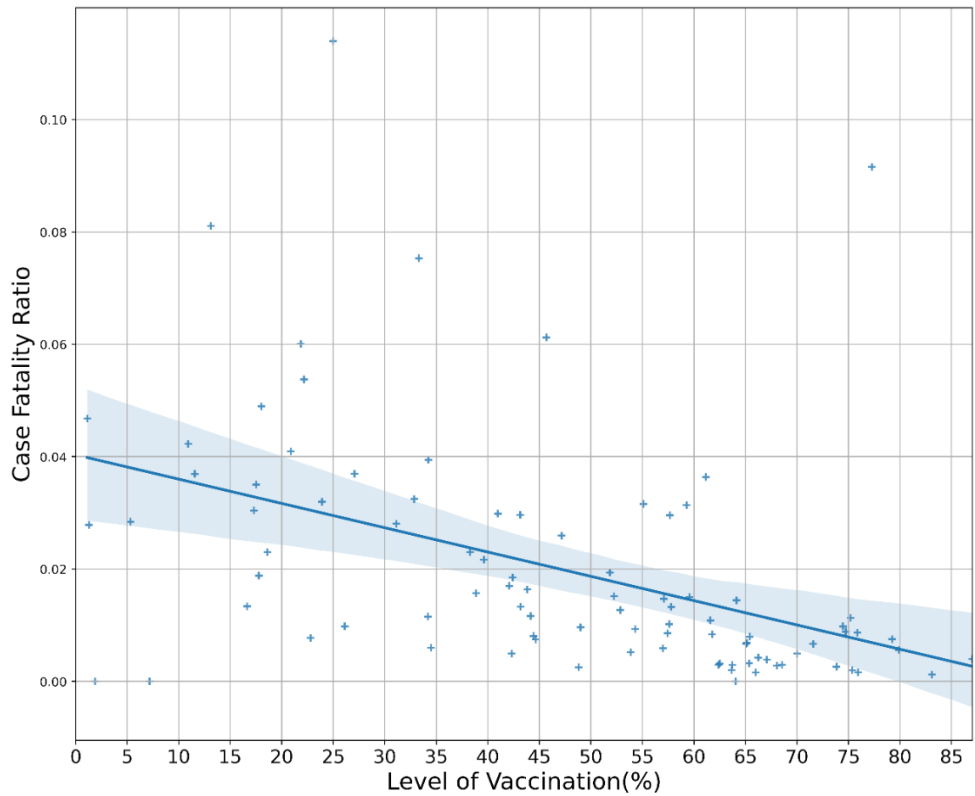

**Fig. 3** Regression plot between vaccination level and CFR

To have the exact value of CFR, the death cases should be a subset of confirmed cases tracked over time [9]. However, because to the lack of such data, we used the total number of cases over the same time period to generate an overall estimate as a simplified form of CFR calculation.

To avoid selection bias of duration, we observed CFR throughout the last three weeks of December 2021, *duration-3*, and took the vaccination level on December 10, 2021 into account. The outcome is shown in Fig. 4. The downward trend of the CFR values for higher vaccinated group was observed here too.

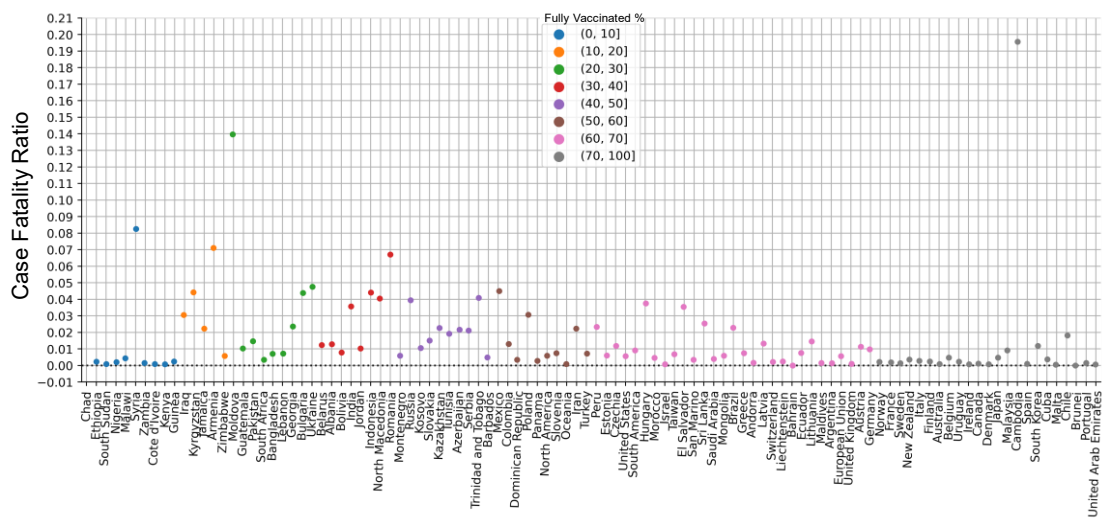

**Fig. 4** *CFR*s of different countries at *duration*-3

According to Fig. 4, the United States, Ecuador, Morocco, Argentina, and numerous other countries with vaccination coverage of more than 60% had *CFR*s of less than 0.01 or 1 percent at duration-3, however they were observed with vaccination coverage of less than 60% and CFRs above 0.01 across *duration-2* (Fig-2). In the vaccine group (70-100], Cambodia was an exception. An in-depth case study of Cambodia's Covid-19 severity may explain the phenomenon.

We also applied linear regression on the data of *duration-3*, the outcome of which is presented in Fig. 5. The vaccination level *coefficient* on CFR was found -0.0004. This also shows the positive impact of vaccination.

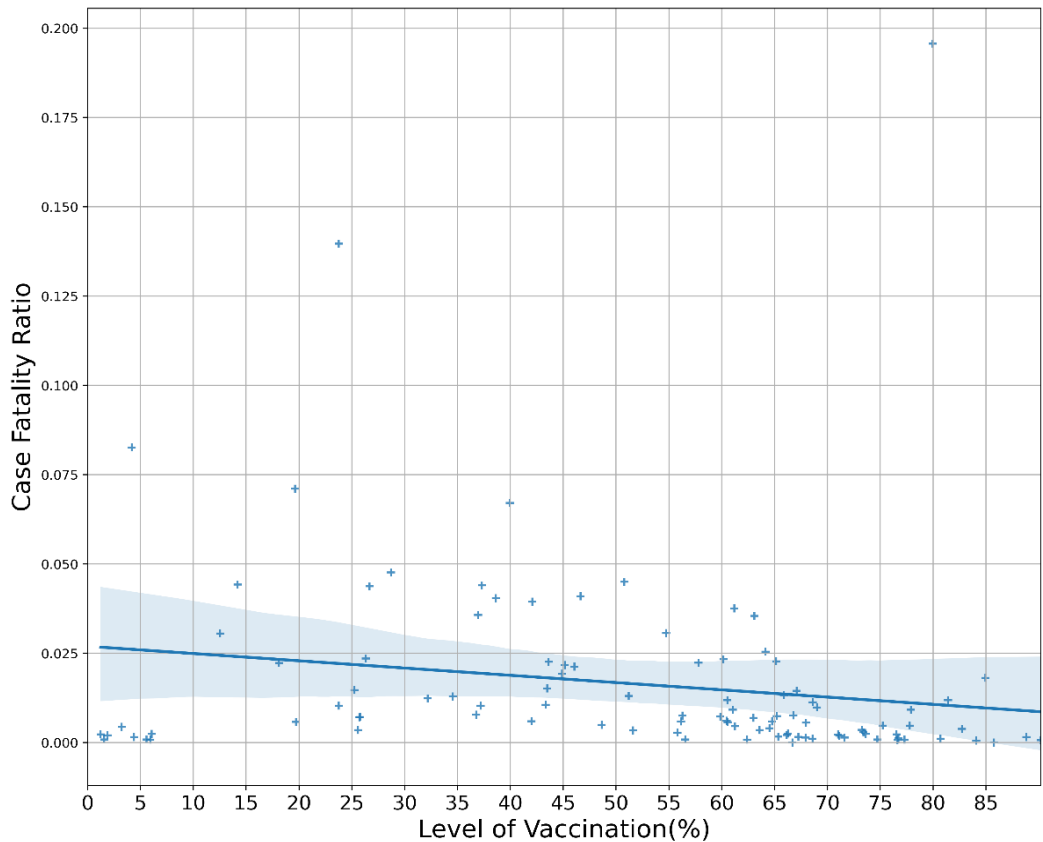

**Fig. 5** Regression plot between vaccination level and CFR (*duration-3*)

Fig. 6 illustrates the mean CFR across *duration-2* and *duration-3* for different vaccination groups. For both durations, the observed mean *CFR*s were closely similar. As a whole, *CFR* is found to decrease with the level of vaccination with an exception for the case of the mean *CFR*s for the groups of vaccination level (0-10] and (70-100]. Such behavior of the data may be due to other factors including immunized by previous infection and expiration of the duration of immunization. Further studies may reveal more insights in this regard.

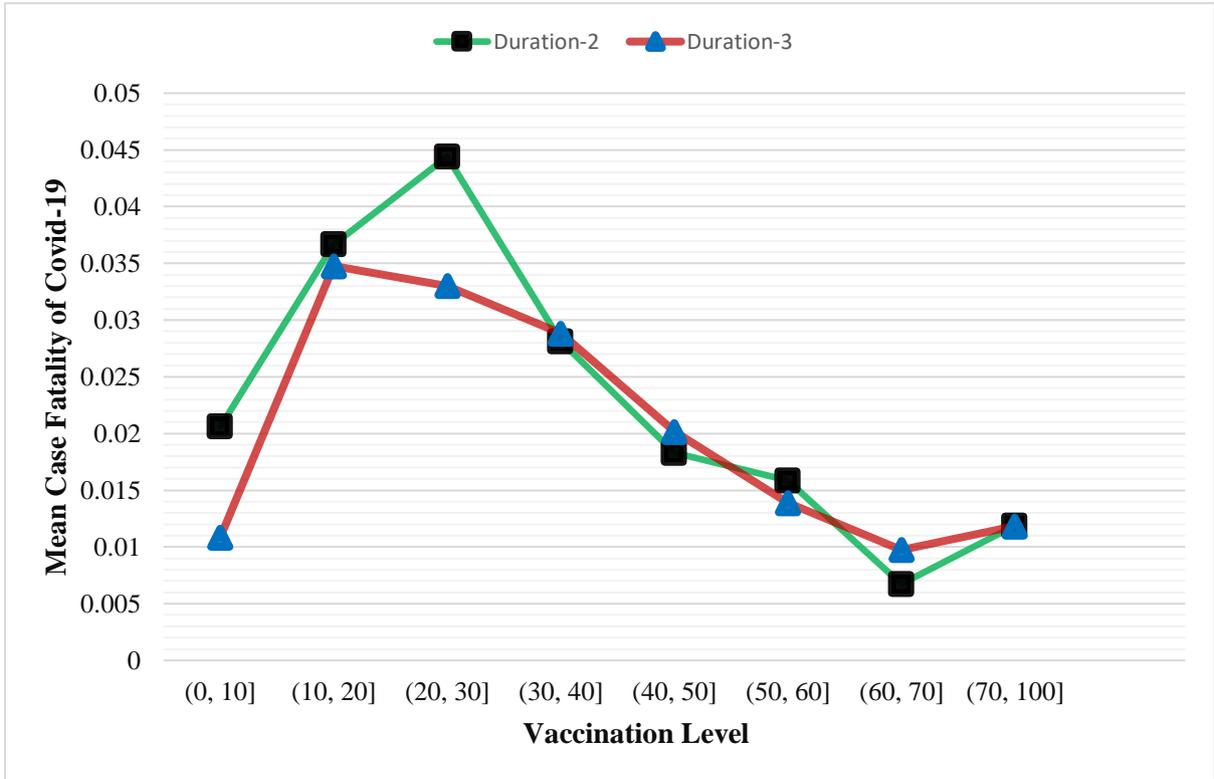

**Fig. 6** Mean *CFR*s for two observed durations

# Discussion

This study examined the effectiveness of the Covid-19 vaccination by examining worldwide epidemiological time series data. Specifically, we suggest that the effectiveness of the SARS-CoV-2 vaccination is better understandable based on the case fatality ratio (*CFR*). This discovery may help to reduce cynicism about vaccine effectiveness.

Vaccinations may do a little to stop the spread of the SARS-CoV-2 virus since no correlation of the vaccination level of country is found with the rate of infections. However, the impact of vaccinating the population is evident in a country's case fatality ratio (*CFR*).

According to the findings of this study, countries with higher vaccination levels will experience less severity of Covid-19 than countries with lower vaccination levels. As a result, the aim for global policymakers should be to make vaccinations available to all countries and to implement policies that would allow the governments to quickly surpass the 65% immunization level.